\documentclass[a4paper]{jpconf}
\usepackage{graphicx}
\begin{document}
\title{Effect of electron-phonon coupling in the ARPES spectra of the tri-layer cuprate Bi$_2$Sr$_2$Ca$_2$Cu$_3$O$_{10+\delta}$}

\author{S.~Ideta$^1$, T.~Yoshida$^1$, M.~Hashimoto$^1$, A.~Fujimori$^1$, H.~Anzai$^2$, A.~Ino$^2$, M.~Arita$^3$, H.~Namatame$^3$, M.~Taniguchi$^{2, 3}$, K.~Takashima$^1$, K.~M.~Kojima$^1$, S.~Uchida$^1$}

\address{$^1$Department of Physics, University of Tokyo, Bunkyo-ku, Tokyo 113-0033, Japan\\
$^2$Graduate School of Science, Hiroshima University, Higashi-Hiroshima 739-8526, Japan\\
$^3$Hiroshima Synchrotron Center, Hiroshima University, Higashi-Hiroshima 739-0046, Japan\\}

\ead{ideta@ap.t.u-tokyo.ac.jp}

\begin{abstract}
Angle-resolved photoemission spectroscopy using tunable low energy photons allows us to study the quasi-particle (QP) dispersions of the inner and outer CuO$_2$ planes (IP and OP) separately in the tri-layer cuprate Bi$_2$Sr$_2$Ca$_2$Cu$_3$O$_{10+\delta}$ (Bi2223). The kink energy of the OP band is $\sim$ 70 meV, as observed in various high-$T_c$ cuprates, while that of the IP band is as large as $\sim$ 100 meV in the superconducting (SC) state. This large kink energy is attributed to the $\sim$35 meV buckling mode plus the large ($\sim$ 60 meV) SC gap of IP. The IP band also shows a weak kink feature at $\sim$ 70 meV in the SC state. The latter feature can be explained either by the 70 meV half-breathing mode or by the $\sim$ 35 meV buckling-phonon mode plus the $\sim$ 40 meV SC gap of OP if interlayer scattering of QP is involved. 
\end{abstract}

\section{Introduction}

\ \ \ \ Attractive interaction between two electrons mediated by a virtual phonon excitation is the origin of Cooper paring in conventional superconductors. Virtual boson excitations are also likely to be the origin of Cooper paring in the high-temperature cuprate superconductors (HTSCs), but the nature of the bosons in HTSCs, phonons or magnetic excitations, has remained controversial. Because electron-boson coupling causes a change in the Fermi velocity below the boson energy, resulting in a kink in the  quasi-particle (QP) dispersion curves. It has been controversial whether the kink is due to phonons \cite{Lanzara,  Cuk1, Yamasaki,HYKee, Cuk2,YChen_kink, JinhoLee} or magnetic excitations \cite{Bogdanov1, HFFong, Carbotte, Dahm, HFFong2}. The kink in the nodal region is relatively simple because of the absence of a $d$-wave superconducting (SC) gap in that direction, and may allow us a relatively straightforward interpretation. According to previous theoretical work \cite{Sandvik, Devereaux}, the nodal QP can be scattered into the antinodal region, where the SC gap is maximum $\sim\Delta$ and, therefore, the nodal kink energy can be given by $E_{\rm{kink}}$ = $\Omega$ + $\Delta$ in the SC state and $E_{\rm{kink}}$ = $\Omega$ in the normal state, where $\Omega$ is the energy of the boson mode and $\Delta$ is the magnitude of the SC order parameter at the anti-node.

ARPES experiment on the optimally doped double-layer Bi$_2$Sr$_2$CaCu$_2$O$_{8+\delta}$ (Bi2212) \cite{Lee2} has shown the nodal kink energy to be $E_{\rm{kink}}$ $\sim$ 70 meV in the SC state, which is interpreted as the sum of the boson mode energy $\Omega\sim$ 35 meV and the anti-nodal SC gap $\Delta\sim$ 35 meV, consistent with the relationship $E_{\rm{kink}}$ = $\Omega$ + $\Delta$. Here, one of the likely candidates for the $\sim$ 35 meV mode is the out-of-plane vibration (buckling mode) of Cu and O atoms. However, an in-plane optical phonon mode (half-breathing mode), which does not strongly scatter the nodal QP into the anti-nodal region but scatters the nodal QP to another nodal region with $\Delta$ = 0 \cite{Devereaux}, should have an energy $\sim$ 70 meV, too, according to $E_{\rm{kink}}$ = $\Omega$+$\Delta$. Therefore, one cannot determine whether the kink at 70 meV is due to the buckling mode which scatters the nodal QP to the antinodal region where the SC gap opens, or due to the half-breathing mode which scatters the nodal QP to another nodal region, or both.  

In order to study the origin of the kink, we have performed an ARPES study on the tri-layer cuprate Bi$_2$Sr$_2$Ca$_2$Cu$_3$O$_{10+\delta}$ (Bi2223), which shows the highest $T_c$ of 110 K among the Bi-based HTSCs. The ARPES spectra of Bi2223 reveal two bands corresponding to the outer and inner CuO$_2$ planes (OP and IP) with different $\Delta$'s for the OP and IP \cite{Ideta_twoband}. We have found that the kink energies in the nodal direction are remarkably different between the IP and OP bands, which can be attributed to the different anti-nodal $\Delta$'s between IP and OP, consistent with the scenario $E_{\rm{kink}}$ = $\Omega$ + $\Delta$ \cite{Sandvik, Devereaux}. Furthermore, we have identified a fine structure in the kink of the IP band, which may be partially attributed to the scattering of the nodal QP from IP to OP through the excitation of a buckling phonon mode.

\section{Experimental}
\ \ \ \ Single crystals of optimally doped Bi2223 ($T_c$ = 110 K) were grown by the traveling solvent floating zone method. ARPES experiments were carried out at beam line 9A of the Hiroshima Synchrotron Radiation Center (HiSOR), Hiroshima University, using circularly polarized light with $h\nu$ = 7.55, 8.4, and 11.95 eV. The total energy resolution ($\Delta E$) was set better than 5 meV. The samples were cleaved $\it{in}$ $\it{situ}$ under an ultrahigh vacuum of $\sim$1$\times$10$^{-11}$ Torr.

\section{Results and Discussion}
\begin{figure}[t]
%\begin{minipage}{14pc}
\begin{center}
\includegraphics[width=27pc]{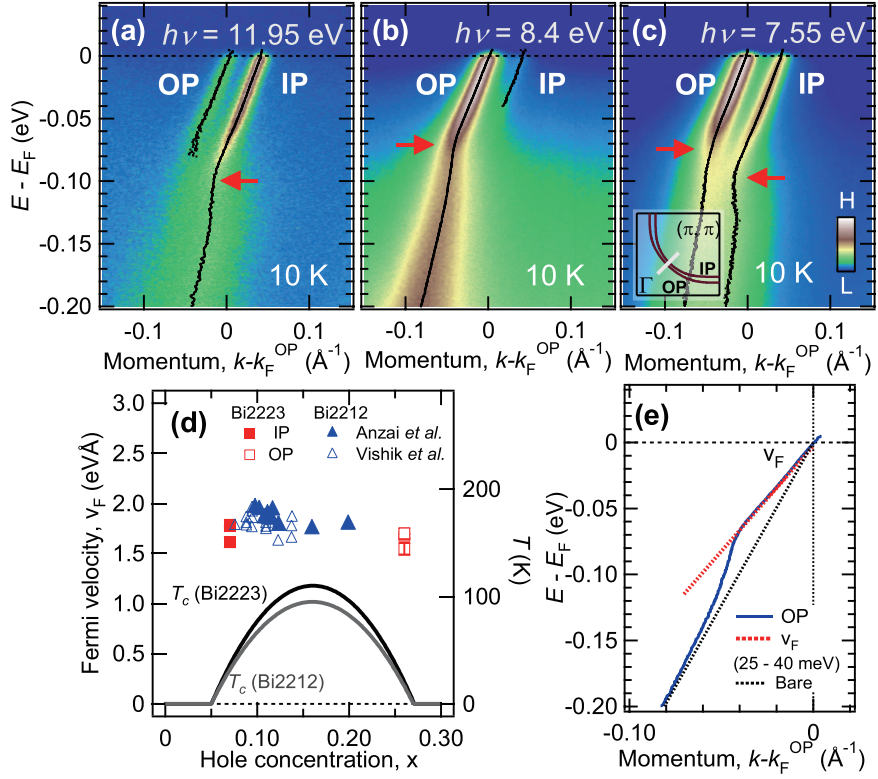}
\end{center}
\caption{\label{label} ARPES spectra of the tri-layer Bi2223 in the nodal direction. (a)-(c): Outer-plane (OP) and inner-plane (IP) bands in the superconducting state taken with different photon energies. Red arrows show an anomaly of the band dispersion, i.e., a kink. (d): Fermi velocities ($v_{\rm{F}}$'s) for the OP and IP bands plotted against hole concentration. The $v_{\rm{F}}$'s of Bi2212 defined by the average group velocity between $E_{\rm{F}}$ and 40 meV below it for various hole concentrations are also shown. (e): Dispersion of the OP band derived from panel (b). Black dotted line denotes the assumed linear bare band. Red dotted line is a linear fit to the dispersion from $E_{\rm{F}}$ to 25 - 40 meV below it. }
\label{Fig1_kink}
%\end{minipage}\hspace{2pc}%
\end{figure}

\ \ \ \ As shown in Figs. \ref{Fig1_kink}(a) - \ref{Fig1_kink}(c), the OP and IP bands are separately observed in the nodal direction by tuning the photon energy. Black dots indicate band dispersions determined by the peak positions of the momentum distribution curves (MDCs). One can see that the OP and IP bands show anomalies at $\sim$ 75 and $\sim$ 100 meV, respectively, as shown by a red arrow in the figures. In Fig. \ref{Fig1_kink}(d), Fermi velocities ($v_{\rm{F}}$'s) for the OP and IP bands are plotted as functions of hole concentration estimated from the Fermi surface area \cite{Ideta_twoband}, and show a constant value. Here, we note that the $v_{\rm{F}}$ of Bi2223 has been estimated by a linear fit from the Fermi level ($E_{\rm{F}}$) to 30 - 40 meV below it. The $v_{\rm{F}}$ of Bi2212 in previous ARPES studies defined by the average group velocity in the same energy range is also plotted \cite{Inna_kink, Anzai}, showing a nearly constant value \cite{Zhou2}.    (The doping-dependent low energy kink observed at $<$ 20 meV \cite{Inna_kink, Anzai} has been ignored in the present analysis.)

\begin{figure}[t]
%\begin{minipage}{14pc}
\begin{center}
\includegraphics[width=38pc]{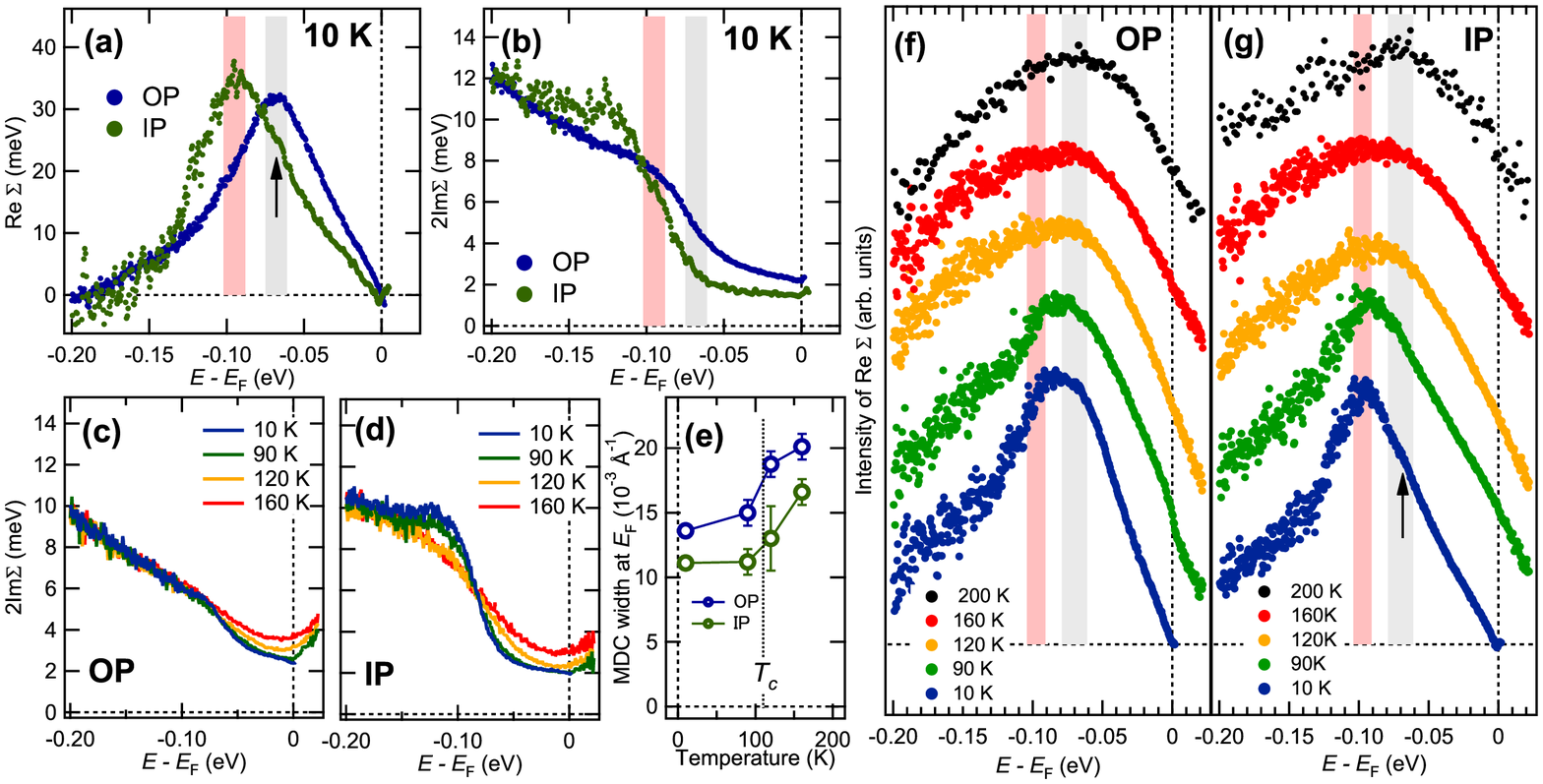}
\end{center}
\caption{\label{label} Real and imaginary parts (Re$\Sigma$ amd Im$\Sigma$) of the self-energy $\Sigma$ of Bi2223. (a),(b): Re$\Sigma$ and 2Im$\Sigma$ of the OP and IP bands. Shaded regions indicate the major kink energies. Re$\Sigma$ has been obtained by assuming a linear bare band shown in Fig. \ref{Fig1_kink}(e). (c), (d): 2Im$\Sigma$ for the OP and IP bands at various temperatures. (e): MDC width at $E_{\rm{F}}$ of the OP and IP bands at various temperatures. (f),(g): Re$\Sigma$ of the OP and IP bands at various temperatures. Gray and pink shaded regions highlight the Re$\Sigma$ peak positions of the OP and IP bands, respectively. }
\label{Fig2_kink}
%\end{minipage}\hspace{2pc}%
\end{figure}

In order to analyze the behaviors of the kinks more quantitatively, the real and imaginary parts of the electron self-energy, Re$\Sigma$ and Im$\Sigma$, have been deduced as shown in Figs. \ref{Fig2_kink}(a) and \ref{Fig2_kink}(b), respectively. Here, Re$\Sigma$ has been deduced from the difference between the observed QP dispersion and the bare band dispersion, which is assumed to be linear as shown in Fig. \ref{Fig1_kink}(e). The peak positions of Re$\Sigma$ corresponding to the energies of the kinks for OP and IP are shown by shaded region. One finds that the peaks of Re$\Sigma$ for the OP and IP bands obviously occur at different energies: for OP $\sim$ 75 meV, similar to that of Bi2212 \cite{Lanzara,  Cuk1, Yamasaki,Bogdanov1}, while for IP as large as $\sim$ 100 meV. From the Kramers-Kronig relation, a kink in Re$\Sigma$ is expected to manifest itself as a step in Im$\Sigma$. In fact, Fig. \ref{Fig2_kink}(b) shows that Im$\Sigma$'s for the OP and IP bands exhibit the steepest slope around the corresponding kink energies.

The different kink energies of the OP and IP bands can be easily understood if the $\sim$ 35 meV boson mode scatters the nodal QP into the anti-nodal region, where the SC gaps of different magnitudes open for the OP and IP bands. According to our previous studies \cite{Ideta_twoband, Ideta_Fermiarc}, the SC gaps in the anti-nodal region are $\Delta$ = 40 meV for the OP band and $\Delta$ = 60 meV for the IP band. Accordingly, if the formula $E_{\rm{kink}}$ = $\Omega$ + $\Delta$ is applied, $E_{\rm{kink}}=$ $\Omega$ + $\Delta\sim$ 35 + 40 $\sim$ 75 ($\pm$ 5) meV for the OP band and $E_{\rm{kink}}$ = $\Omega$ + $\Delta\sim$ 35 + 60 $\sim$ 95 ($\pm$ 5) meV for the IP band, consistent with experiment, and the unusually large kink energy for the IP band can be understood. Since phonon bands are located from 12 to 72 meV according to optical measurements of Bi2223 \cite{Boris}, the kink energy of $\sim$100 meV observed for the IP band cannot be explained without the contribution of the large SC gap of the IP band. The half-breathing mode of energy $\sim$70 meV will also contribute to a kink at $\sim$70 - 75 meV for both the OP and IP bands though $E_{\rm{kink}}$ = $\Omega + \Delta$ $\sim$ 70 + 0 meV, since the half-breathing mode scatters a nodal QP to another  nodal region, where $\Delta\sim$ 0. Indeed, the weak shoulder at $\sim$70 meV in the Re$\Sigma$ of IP in the SC state may be attributed to the half-breathing mode. Alternatively, the 70 meV shoulder may be attributed to scattering of the nodal QP into the antinodal region of OP through the $\sim$ 35 meV buckling mode.

Figures \ref{Fig2_kink}(c) - \ref{Fig2_kink}(e) show the Im$\Sigma$ and the MDC width at $E_{\rm{F}}$ of the OP and IP bands at various temperatures. Panel (e) shows that the MDC widths at $E_{\rm{F}}$ of the OP and IP bands increase with temperature and shows a discontinuous change around $T_c$ = 110 K, like the previous experimental results on Bi2212 \cite{Yamasaki}. This means that the scattering of the nodal QP is suppressed by the opening of the SC gap. Figures \ref{Fig2_kink}(f) and \ref{Fig2_kink}(g) show Re$\Sigma$ for the OP and IP bands at various temperatures. The peak positions of Re$\Sigma$ do not show an abrupt change at $T_c$ = 110 K but at $T$ = 200 K, the peak position for the IP band becomes $\sim$ 70 meV. Because the SC gap closes at $T_c$ = 110 K and the pseudogap closes at temperature $T^*$ well above $T_c$, the change in the kink energy of the IP band would be attributed to the closure or the reduction of the pseudogap in the antinodal region. The kink energy of $\sim$70 meV at $T\sim$ 200 K is then attributed to the energy of the breathing mode.

As mentioned above, there are two possible explanations for the weak $\sim$70 meV shoulder in the Re$\Sigma$ of IP: (i) the nodal QP of IP is scattered into the other nodal region through the excitation of the half-breathing mode of energy $\sim$ 70 meV, or (ii) the nodal QP  of IP is scattered into the antinodal region of OP through the excitation of $\sim$ 35 meV phonons. If the latter scattering exist, a Cooper pair near the node of IP will be scattered into the antinodal region of OP with the virtual excitation of the $\sim$ 35 meV boson. This can become a mechanism which enhances $T_c$ in multilayer cuprates \cite{Chakravarty}. As the 35 meV boson, we consider the buckling mode \cite{Kovaleva} because such a mode scatters the nodal QP to the anti-nodal one according to Refs. \cite{Devereaux, TPDevereaux_kink2}. 

In conclusion, we have performed an ARPES experiment on Bi2223 and found that the kink energy shows different values between IP and OP: $\sim$ 75 meV for the OP band and $\sim$ 100 meV for the IP band. This difference is attributed to the different SC gap magnitudes between OP and IP. Furthermore, we find that the Re$\Sigma$ of IP shows a shoulder structure at $\sim$ 70 meV, which may indicate that (1) the QP of the IP is scattered by the half-breathing mode, or (2) the inter-layer scattering mediated by the buckling mode occurs.

\ack{\ \ \ \ We would like to thank Z.-X. Shen for enlightening discussion. ARPES experiments were carried out at HiSOR (Proposal No. 07-A-10, and 08-A-35). This work was supported by a Grant-in-Aid for Scientific Research, Award for Young Scientist (B), JSPS, Japan, and  A3 Foresight Program \textquotedblleft Joint Research on Novel Properties of Complex Oxides\textquotedblright. S.I. acknowledges support from the
Japan Society for the Promotion of Science.}

\end{document}